\documentclass[prl,amsmath,amssymb,twocolumn,superscriptaddress]{revtex4-1}
\usepackage{amsmath,amssymb}
\usepackage[usenames]{color}
\usepackage{amssymb}
\usepackage{grffile}
\usepackage[pdftex]{graphicx}
\usepackage{amsmath, amstext, amssymb, amsfonts, amsxtra}
\usepackage{textcomp}
\usepackage{xspace}
\usepackage{bbm}
\usepackage{bm}
\newcommand{\be}{\begin{equation}}
\newcommand{\ee}{\end{equation}}

\newcommand{\como}{Center for Nonlinear and Complex Systems, Dipartimento
di Scienza e Alta Tecnologia, Universit\`a degli Studi dell'Insubria,
via Valleggio 11, 22100 Como, Italy}
\newcommand{\infn}{Istituto Nazionale di Fisica Nucleare, Sezione di Milano,
via Celoria 16, 20133 Milano, Italy}
\newcommand{\NEST}{NEST, Istituto Nanoscienze-CNR, I-56126 Pisa, Italy}
\newcommand{\brazil}{International Institute of Physics, Federal University
of Rio Grande do Norte, Campus Universit\'ario - Lagoa Nova, CP. 1613,
Natal, Rio Grande Do Norte 59078-970, Brazil}
\newcommand{\xiamen}{Department of Physics and Key Laboratory of Low
Dimensional Condensed Matter Physics (Department of Education of Fujian
Province), Xiamen University, Xiamen 361005, Fujian, China}

\begin{document}

\title{Power-efficiency-fluctuations trade-off in steady-state heat engines:
\\ The role of interactions}

\author{Giuliano Benenti}
\affiliation{\como}
\affiliation{\infn}
\affiliation{\NEST}
\author{Giulio Casati}
\affiliation{\como}
\affiliation{\brazil}
\author{Jiao Wang}
\affiliation{\xiamen}

\begin{abstract}
We consider the quality factor $\mathcal{Q}$, which quantifies the
trade-off between power, efficiency, and fluctuations in steady-state
heat engines modeled by dynamical systems. 
We show that the nonlinear scattering theory, both in classical
and quantum mechanics, sets the bound $\mathcal{Q}=3/8$ when approaching
the Carnot efficiency. On the other hand, interacting, nonintegrable and
momentum-conserving systems can achieve the value $\mathcal{Q}=1/2$, which
is the universal upper bound in linear response.
This result shows that interactions are necessary to achieve the optimal 
performance of a steady-state heat engine. 
\end{abstract}

\maketitle

\emph{Introduction.-} Understanding the bounds that a heat engine must
obey is of key importance both for basic science and for technological
development. Ideally, a heat engine should work with \emph{efficiency}
$\eta$ close to the Carnot efficiency $\eta_C$, deliver large \emph{power}
$P$, and have small power \emph{fluctuations} $\Delta_P$. The Carnot limit
is intuitively associated with infinitely slow engines, so that the output
power vanishes. On the other hand, the second law of thermodynamics by
itself does not forbid an engine operating at the Carnot efficiency with
a finite output power~\cite{Benenti2011}. Such a dream engine was denied
in models with inelastic scattering~\cite{Saito2011, Sanchez2011, Horvat2012,
Vinitha2013, Brandner2013a, Brandner2013b, Brandner2015, Yamamoto2016} and
for two-terminal systems on the basis of symmetry considerations for the
Onsager kinetic coefficients~\cite{Luo2020}. 
Moreover, for systems described as Markov
processes, the bound $P\le A(\eta_C-\eta)$ was proven~\cite{Saito2016}, with
$\eta_C=1-T_R/T_L$, $T_L$ and $T_R$ ($T_L>T_R$) being the temperatures of
the hot and the cold reservoir, respectively, and $A$ a system-specific
constant. On the other hand, $A$ may diverge when approaching the Carnot
efficiency~\cite{Allahverdyan2013, Shiraishi2015, Campisi2016, Koning2016,
Polettini2017,Lee2017}, 
for instance, when the engine working fluid is at the verge
of a phase transition, and therefore the Carnot efficiency may be approached
at finite power. However, fluctuations make impractical such
engines~\cite{Holubev2017}.

On the base of thermodynamic uncertainty relations~\cite{Barato2015,
Gingrich2016, Seifert2019, Horowitz2020} for the work current (i.e., for
the power delivered by the engine), a trade-off encompassing efficiency,
power, and fluctuations has been proven by Pietzonka and
Seifert~\cite{Pietzonka2018}, for a large class of 
steady-state classical stochastic heat engines with time-reversal symmetry. 
Such class includes engines with a discrete set of internal
states described by thermodynamically consistent rate equations, and
continuous systems modeled with an overdamped Langevin dynamics. 
The bound reads
\be
\mathcal{Q}\equiv P\frac{\eta}{\eta_C-\eta}\frac{k_B T_R}{\Delta_P}\le\frac{1}{2},
\label{eq:seifertbound}
\ee
where $k_B$ is the Boltzmann constant and the power fluctuations are
measured by~\cite{footnote_DeltaP}
\be
\Delta_P=\lim_{t\to\infty} [P(t)-P]^2 t,
\label{eq:DeltaP}
\ee
where $P(t)$ is the mean delivered power up to time $t$.

In this paper, at difference from the above stochastic thermodynamics approach, 
we examine for \emph{purely dynamical models} the 
upper bound to the quality factor $\mathcal{Q}$.
We focus on the most desirable regime for a heat engine, i.e., 
\emph{when approaching the Carnot efficiency} 
at the largest possible output power.  
We first find a general solution to this problem for systems
that can be modeled by the nonlinear scattering
theory. That is, for noninteracting
systems or more generally for systems in which interactions can be treated
at a mean-field Hartree level. In this case we prove that 
the quality factor $\mathcal{Q}<3/8$, and that the limit 
value $\mathcal{Q}=3/8<1/2$
is only achieved for $\eta\to\eta_C$. Scattering theory sets therefore 
a stronger bound to the quality factor $\mathcal{Q}$ 
than Eq.~(\ref{eq:seifertbound}).
We stress that the above results are valid both in classical and 
in quantum mechanics.
We then consider the class of interacting,  
non-integrable momentum-conserving
systems. This is, to our knowledge, the only class of interacting 
dynamical systems which is known to achieve, at the thermodynamic limit,
 the Carnot efficiency~\cite{Benenti2013, Benenti2014, Chen2015}.
We show in a concrete example of a nonintegrable gas of 
elastically colliding particles that such systems 
saturate the bound $\mathcal{Q}=1/2$. This value
is achieved in the tight-coupling limit, where the Onsager matrix of
kinetic coefficients becomes singular.


\emph{Bound from scattering theory.-}
For concreteness, hereafter we consider thermoelectric transport, even
though our results could be equally applied to other examples of
steady-state conversion of heat to work, like thermodiffusion. In the
Landauer-B\"uttiker quantum scattering theory, the electrical current,
flowing from the left to the right reservoir, reads
\be
J_e=\frac{e}{h}\int_{-\infty}^{\infty} d\epsilon \,{\cal T}(\epsilon)
\,[f_L(\epsilon)-f_R(\epsilon)],
\ee
where $e$ is the electron charge, $h$ the Planck constant,
${\cal T}(\epsilon)$ the transmission probability for a particle with
energy $\epsilon$ to transit from one end to another of the system
($0\le {\cal T}(\epsilon)\le 1$), and $f_\alpha(\epsilon)=\{1+\exp[
(\epsilon-\mu_\alpha)/k_B T_\alpha]\}^{-1}$ is the Fermi distribution
function for reservoir $\alpha$ ($\alpha=L,R$), at temperature $T_\alpha$
and electrochemical potential $\mu_\alpha$~\cite{footnote_modes}.
The heat current that flows into the system from reservoir $\alpha$ is
\be
J_{h,\alpha}=\frac{1}{h}\int_{-\infty}^\infty d\epsilon\, (\epsilon-\mu_\alpha)
\,{\cal T}(\epsilon)
\,[f_L(\epsilon)-f_R(\epsilon)].
\ee

The output power $P=(\Delta V) J_e$, where $\Delta V=\Delta \mu/e$ is the
applied voltage, with $\Delta\mu=\mu_R-\mu_L>0$. The efficiency of heat to
work conversion is given by $\eta=P/J_{h,L}$, with $P,J_{h,L}>0$.
The transmission function which maximizes the efficiency for a given power
is a boxcar function, ${\cal T} (\epsilon)=1$ for
$\epsilon_0<\epsilon<\epsilon_1$ and ${\cal T}(\epsilon)=0$
otherwise~\cite{Whitney2014,Whitney2015}. Here $\epsilon_0=\Delta\mu/\eta_C$
is obtained from the condition $f_L(\epsilon_0)=f_R(\epsilon_0)$ and
$\epsilon_1$ can be determined numerically by solving the equation
$\epsilon_1=\Delta\mu J_{h,L}^\prime/P^\prime$, where the prime indicates
the derivative over $\Delta\mu$ for fixed ${\cal T}$ (this equation is
transcendental since $J_{h,L}$ and $P$ depend on $\epsilon_1$). The maximum
achievable power $P_{\rm max}$ is obtained when $\epsilon_1\to\infty$
and is given by $P_{\rm max}\approx 0.317\,k_B^2 (\Delta T)^2/h$, with
$\Delta T=T_L-T_R>0$. In the limit $\epsilon_1\to\epsilon_0$, known
as delta-energy filtering~\cite{mahan, linke1, linke2}, $P\to 0$ and
$\eta\to\eta_C$.

\begin{figure}[!b]
\includegraphics[width=8.9cm]{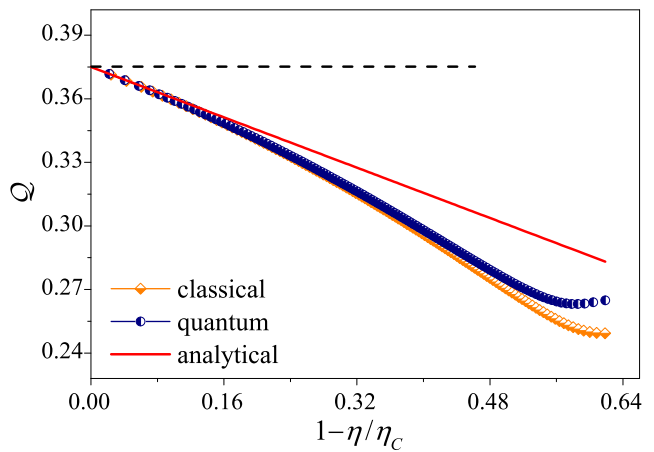}
\caption{Quantum and classical quality factor $\mathcal{Q}$ based on the
scattering theory. For $\eta$ close to $\eta_C$, numerical results are in
excellent agreement with the analytical formula (\ref{eq:expandQ}) up to
the linear term (red solid line). The dashed line indicates the bound
$\mathcal{Q}=3/8$ 
for reference. Parameter values adopted in simulations are $T_L=1$,
$T_R=0.9$, $\mu_L=0$. For any value of power, $\mu_R=\Delta \mu$ is determined
in the optimization procedure described in the text. Here we use units such
that $k_B=e=h=1$.}
\label{fig:Qnoninteracting}
\end{figure}

The power fluctuations can be computed from the Levitov-Lesovik cumulant
generating function~\cite{Levitov1993, Nazarov2009, Segal2019}. For the
above boxcar function, we obtain
\be
\Delta_P=\frac{(\Delta\mu)^2}{h}\int_{\epsilon_0}^{\epsilon_1}
 d\epsilon
\,[f_L(\epsilon)+f_R(\epsilon)-f_L^2(\epsilon)-f_R^2(\epsilon)].
\ee


Using the above defined expressions for $\eta$, $P$, and $\Delta_P$, 
we can compute the quality factor $\mathcal{Q}$.
As detailed in the supplemental material~\cite{supp}, we can
expand $\mathcal{Q}$ close to the Carnot efficiency, that is,
for $1-\eta/\eta_C\ll 1$. We then obtain the analytical result
\be
\mathcal{Q}=\frac{3}{8}-
\frac{9}{128}\frac{T_L+T_R}{T_R}\left(1-\frac{\eta}{\eta_C}\right)+
\mathcal{O}\left[\left(1-\frac{\eta}{\eta_C}\right)^2\right].
\label{eq:expandQ}
\ee
When going far from the Carnot limit, the dependence of $\mathcal{Q}$ 
on efficiency can be computed numerically. The results are shown in  
Fig.~\ref{fig:Qnoninteracting}, for the optimal boxcar function, which
maximizes efficiency for any value of power. We can see that
$\mathcal{Q}<3/8$ for any value of $\eta$, the value $\mathcal{Q}=3/8$
being obtained only for $\eta=\eta_C$ (correspondingly, $P=0$). 
A similar analysis can be performed in the classical case~\cite{supp}, and
for the optimal boxcar function~\cite{Luo2018} expansion (\ref{eq:expandQ})
is still valid. As shown in Fig.~\ref{fig:Qnoninteracting}, classical and
quantum $\mathcal{Q}$ differ at higher orders, with the quantum quality
factor slightly larger than the classical one.

\emph{Momentum-conserving systems.-}
The above results raise two interesting questions:  
(i) Is it possible to find interacting
systems which may overcome the scattering theory bound $\mathcal{Q}=3/8$?
(ii) Is it possible to approach or even overcome bound (\ref{eq:seifertbound}),
$\mathcal{Q}=1/2$, when approaching the Carnot efficiency? 
To address
this question, we consider non-integrable momentum-conserving
systems, for which the Carnot limit 
can be achieved at the thermodynami 
limit~\cite{Benenti2013, Benenti2014, Chen2015}. 
We perform nonequilibrium
simulations, with the momentum-conserving system (specifically, a classical
one-dimensional diatomic gas of elastically colliding
particles~\cite{Benenti2013}) in contact with two reservoirs at different
temperatures and electrochemical potentials, which maintain stationary
coupled energy and particle flows. In our simulations, particles are
absorbed whenever they hit a reservoir, while the two reservoirs inject
particles with rates and energy distributions determined by their
temperatures and electrochemical potentials~\cite{reservoir}.

\begin{figure}[!]
\includegraphics[width=8.9cm]{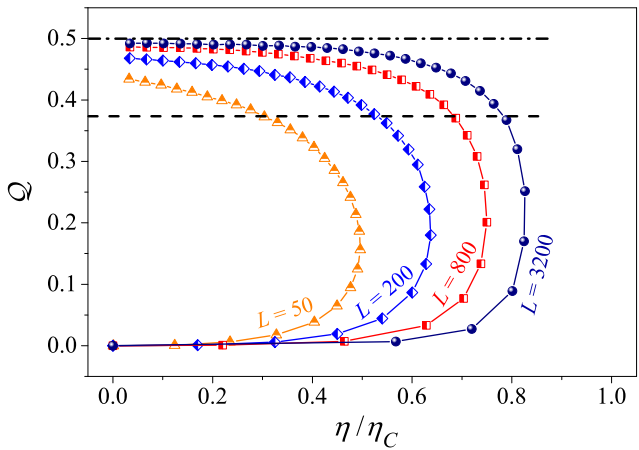}
\caption{Quality factor versus efficiency for a one-dimensional diatomic
gas of elastically colliding particles of masses $m=1$ and $M=3$ with
$T_L=1.05$, $T_R=0.95$, $\mu_L=-\Delta \mu/2$, and $\mu_R=\Delta \mu/2$.
The system size $L$ is equal to the mean number of particles inside the
system. The dashed and the dot-dashed line indicate the bound 
$\mathcal{Q}=3/8$
and $\mathcal{Q}=1/2$, respectively. We set $k_B=e=1$.
}
\label{fig:Qinteracting}
\end{figure}

From our numerical simulations, we can determine the charge and heat
currents, and consequently power, efficiency, and fluctuations. In
Fig.~\ref{fig:Qinteracting}, we compute the trade-off $\mathcal{Q}$ for
different system sizes $L$. Note that these curves have two values for
a given value of efficiency $\eta$, as they are obtained by changing
$\Delta \mu$ from zero, where $\eta=0$, up to the stopping value, where
the electrochemical potential difference becomes too high to be overcome
by the temperature difference, and again $\eta=0$. The branch with higher
values of $\mathcal{Q}$ corresponds to the lower values of $\Delta\mu$.
Note that the scattering theory bound $\mathcal{Q}=3/8$ is overcome,
up to higher values of the efficiency as the system size increases.
At the same time, bound (\ref{eq:seifertbound}), $\mathcal{Q}=1/2$,
is approached closer and closer.

\emph{Linear response results.-}
For a given temperature difference
$\Delta T=T_L-T_R$, the gradient $|\nabla T|=\Delta T/L$ decreases with
the system size $L$. We can then apply the linear response theory
in the large-$L$ regime, which is the most interesting one in 
our model, as the Carnot efficiency is achieved when $L\to\infty$.
Within linear response, the charge and heat
currents (from the left to the right reservoir) are given by~\cite{Callen,
Groot, Benenti2017}
\begin{equation}
\left(
\begin{array}{c}
J_e\\
J_h
\end{array}
\right) = \left(
\begin{array}{cc}
L_{e e} & L_{e h} \\
L_{h e} & L_{h h}
\end{array}
\right) \left(
\begin{array}{c}
\mathcal{F}_e\\
\mathcal{F}_h
\end{array}
\right) ,
\label{eq:lresponse}
\end{equation}
where $\mathcal{F}_e=-\nabla [\mu /(ek_B T)]$ and
$\mathcal{F}_h=\nabla [1/(k_BT)]$
are the thermodynamic forces and $L_{ij}$ ($i,j=e,h$) the Onsager kinetic
coefficients, which obey, for systems with time-reversal symmetry, the
Onsager reciprocal relation $L_{eh}=L_{he}$. The matrix of the kinetic
coefficients is known as the Onsager matrix $\mathbb{L}$. The second law
of thermodynamics imposes $L_{ee}\ge 0$, $L_{hh}\ge 0$, and
$\det \mathbb{L}\ge 0$.

We consider a generic linear combination of the currents,
$J_{\bf c}=\sum_\alpha c_\alpha J_\alpha={\bf c}^T \mathbb{L}
{\boldsymbol{\mathcal{F}}}$, where ${\boldsymbol{\mathcal{F}}}$ is the
vector of thermodynamic forces. Thermodynamic uncertainty relations are
saturated when ${\bf c}\parallel {\boldsymbol{\mathcal{F}}}$ on the
orthogonal complement of the kernel of $\mathbb{L}$~\cite{Macieszczak2018}.
In nonintegrable momentum-conserving systems, $\mathbb{L}$ becomes
singular in the thermodynamic limit, where the Onsager matrix becomes
singular (a condition known as tight-coupling limit, for which the Carnot
efficiency is achievable). In this limit, the orthogonal complement of the
kernel of $\mathbb{L}$ is one-dimensional, and therefore the thermodynamic
uncertainty relations are saturated for all ${\bf c}$. In particular, bound
(\ref{eq:seifertbound}) is saturated. In contrast, for finite system sizes
the Onsager matrix is positive and the only current such that
${\bf c}\parallel{\boldsymbol{\mathcal{F}}}$ is the entropy production rate
$\dot{S}=\mathcal{F}_e J_e+\mathcal{F}_h J_h$. Since power fluctuations
$\Delta_P$ are proportional to charge fluctuations and not to $\dot{S}$,
it follows that bound (\ref{eq:seifertbound}) is saturated only at the 
thermodynamic limit.

\begin{figure}[!]
\includegraphics[width=8.9cm]{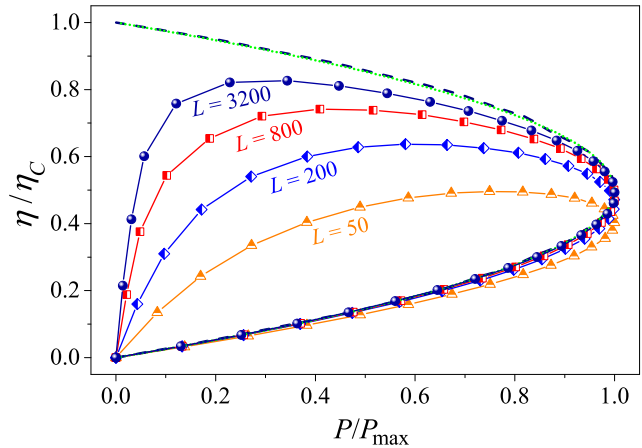}
\caption{The same as Fig.~2 but for efficiency versus power for the diatomic
gas model. The linear response prediction (\ref{eq:LRbound}) in the tight-coupling
limit (green dotted curve) and the upper bound (\ref{eq:seifertbound2})
(blue dashed curve) are also shown.}
\label{fig:Qlinearresponse}
\end{figure}

The validity of using linear response to interpret our
numerical data is numerically confirmed by Fig.~\ref{fig:Qlinearresponse}. For
different system sizes, $\eta/\eta_C$ is shown as a function of
$P/P_{\rm max}$, where $P_{\rm max}$ is the maximum power obtained when
$\Delta \mu$ is varied from zero, where $P=0$, up to the stopping value,
where again $P=0$. In the same figure, we also show the linear response
result in the tight-coupling limit~\cite{Benenti2017}:
\be
\frac{\eta}{\eta_C}=\frac{P/P_{\rm max}}{2(1\pm \sqrt{1-P/P_{\rm max}})}.
\label{eq:LRbound}
\ee
Moreover, we rewrite bound (\ref{eq:seifertbound}) as
\be
\frac{\eta}{\eta_C}\le \frac{1}{1+2P k_B T_R/\Delta_P}
\label{eq:seifertbound2}
\ee
and in the same figure we show the right-hand side of this inequality
for the numerically computed values of power and fluctuations
at the largest available system size, $L=3200$. 
Numerical results, however, suggest
that the upper bound (\ref{eq:seifertbound2}) does not depend on the system
size. The tendency to saturate this bound when increasing the system size
is clearly seen in Fig.~\ref{fig:Qlinearresponse}.
The excellent agreement between
linear response expectations in the tight-coupling limit and the numerically
computed upper bound on efficiency for given power and fluctuations, shows
that linear response theory provides a satisfactory explanation of our
results.

\emph{Discussion and conclusions.-}
We have shown that for steady-state heat engines the
power-efficiency-fluctuations trade-off $\mathcal{Q}\le 3/8$ within
the scattering theory, 
the upper bound $\mathcal{Q}=3/8$ being saturated
at the Carnot efficiency. 
These conclusions hold both in classical and in quantum mechanics.
On the other hand, interacting nonintegrable
momentum-conserving systems may overcome this limit and saturate the
linear-response upper bound $\mathcal{Q}=1/2$. This value is obtained
in the tight-coupling limit, when the Onsager matrix becomes singular
and the Carnot limit can be achieved. From the viewpoint of thermoelectric
transport, our results confirm the relevance of the figure of merit
$ZT=L_{eh}^2/\det \mathbb{L}$~\cite{Benenti2017}: Not only 
the Carnot efficiency can be achieved at the tight-coupling limit 
$ZT\to\infty$, but also the power-efficiency-fluctuations bound 
$\mathcal{Q}=1/2$ is saturated in the same limit. The trade-off 
$\mathcal{Q}$ could be investigated experimentally in the context 
of cold atoms, where a thermoelectric heat  engine with high $ZT$ 
has already been demonstrated, both for weakly~\cite{Brantut2013} 
and strongly interacting particles~\cite{Husmann2018}.

Our analysis does not include the effects of a magnetic field. However,
for two-terminal systems our conclusions would not change. Indeed, the
scattering-theory bound discussed in this paper is the same, irrespective
of whether time-reversal symmetry is broken by an external magnetic field
or not. Moreover, the Onsager matrix obeys the reciprocal relation
$L_{eh}=L_{he}$ for interacting systems, even 
in presence of a generic magnetic
field~\cite{Luo2020}. The effects of a magnetic field on thermoelectric
efficiency were investigated in systems with three or more terminals, which
mimic inelastic scattering events, but in that case it was not possible to
achieve the Carnot efficiency~\cite{Horvat2012, Vinitha2013, Brandner2013a,
Brandner2013b, Brandner2015, Yamamoto2016}. It remains therefore as an
interesting question whether the linear-response bound $\mathcal{Q}=1/2$
may be overcome by a, classical or quantum, interacting model when approaching
the Carnot limit. More generally, we wonder whether stringent bounds from
the scattering theory also apply for periodically driven systems.

We acknowledge support by the NSFC (Grant No. 11535011)
and by the INFN through the project QUANTUM.

\appendix 

\section{Supplemental Material}

Here we provide more details on the derivation of the quality factor
$\mathcal{Q}$ from the scattering theory, for efficiency close to the Carnot 
efficiency, Eq.~(6) of the main text. Hereafter we set $k_B=e=h=1$. In the 
limit $\delta\equiv \epsilon_1-\epsilon_0\to 0^+$, the power $P\sim \delta^2$, 
the deviation from the Carnot efficiency $\eta_C-\eta\sim \delta$, the 
fluctuations $\Delta_P\sim\delta$, so that the trade-off factor $\mathcal{Q}$ 
goes to a constant when $\delta\to 0^+$. More precisely, for the power we 
obtain
\be
P=(\Delta V)\left[{\rm sech}\left(\frac{\Delta V}{2 \Delta T}\right)\right]^2
\frac{\Delta T}{8 T_L T_R}\,\delta^2+\mathcal{O}(\delta^3),
\ee
where in the limit $\delta\to 0^+$ the voltage
$\Delta V=\alpha \Delta T$~\cite{Whitney2015}, with $\alpha\approx 3.24$,
the root of the transcendental equation
\be
\alpha\tanh \left(\frac{\alpha}{2}\right)=3.
\ee
We can then rewrite the power as
\be
P=\frac{\alpha (\Delta T)^2}{4(1+\cosh \alpha)T_L T_R}\,\delta^2
+\mathcal{O}(\delta^3).
\ee
Similarly, we obtain the heat current that flows from the hot reservoir,
\be
J_{h,L}=\frac{\alpha \Delta T}{4(1+\cosh \alpha) T_R}\,\delta^2,
+\mathcal{O}(\delta^3),
\ee
and the efficiency
\be
\eta=\eta_C-\frac{2\eta_C}{3\alpha T_L}\,\delta +
\mathcal{O}(\delta^2).
\ee
To compute power fluctuations, we use the formula~\cite{Segal2019}
$$
\Delta_P=(\Delta V)^2\int_{-\infty}^{+\infty} d\epsilon
({\cal T}(\epsilon)\{f_L(\epsilon)[1-f_L(\epsilon)]
$$
\be
+f_R(\epsilon)[1-f_R(\epsilon)]\}+
{\cal T}(\epsilon)[1-{\cal T}(\epsilon)]
[f_L(\epsilon)-f_R(\epsilon)]^2),
\ee
which reduces to Eq.~(5) of the main paper for the boxcar transmission
function we are considering. Upon expansion for small $\delta$, we obtain
\be
\Delta_P=\frac{\alpha^2(\Delta T)^2}{1+\cosh\alpha}\,\delta
+\mathcal{O}(\delta^2).
\ee
Finally, we derive
\be
\mathcal{Q}=\frac{3}{8}-
\frac{T_L+T_R}{64 T_LT_R}\frac{3}{\alpha}\,\delta+
\mathcal{O}(\delta^2),
\ee
which reduces to Eq.~(6) of the main text after inverting the relation
between efficiency and width of the transmission window:
\be
\delta=\frac{3}{2}\,\alpha T_L \left(1-\frac{\eta}{\eta_C}\right)+
\mathcal{O}\left[\left(1-\frac{\eta}{\eta_C}\right)^2\right].
\ee

In the classical case, the Boltzmann distribution function
$f_\alpha^{(B)}(\epsilon)=\exp[-(\epsilon-\mu_\alpha)/T_\alpha]$
($\alpha=L,R$) is considered rather than the Fermi distribution function
in the calculation of $J_e$ and $J_{h,\alpha}$~\cite{Saito2010,Luo2018}.
Moreover, the power fluctuations are given by~\cite{Gaspard2013,Brandner2018}
\be
\Delta_P=(\Delta V)^2
\int_{-\infty}^{\infty} d\epsilon \,{\cal T}(\epsilon)
\,[f_L^{(B)}(\epsilon)+f_R^{(B)}(\epsilon)].
\ee
We obtain that 
\be
P=\frac{3 (\Delta T)^2}{2 {\tilde e}^3 T_L T_R}\,\delta^2
+\mathcal{O}(\delta^3),
\ee
\be
\eta=\eta_C-\frac{2\eta_C}{9 T_L}\,\delta +
\mathcal{O}(\delta^2),
\ee
\be
\Delta_P=\frac{18(\Delta T)^2}{{\tilde e}^3}\,\delta
+\mathcal{O}(\delta^2),
\ee
where $\tilde e$ is Euler's number. Finally, we have
\be
\mathcal{Q}=\frac{3}{8}-
\frac{T_L+T_R}{64 T_LT_R}\,\delta+
\mathcal{O}(\delta^2),
\ee
from which we recover Eq.~(6) of the main text after inverting the relation
between efficiency and width of the transmission window:
\be
\delta=\frac{9}{2}\, T_L
\left(1-\frac{\eta}{\eta_C}\right)
+\mathcal{O}\left[\left(1-\frac{\eta}{\eta_C}\right)^2\right].
\ee



\begin{thebibliography}{100}

\bibitem{Benenti2011}
G. Benenti, K. Saito, and G. Casati,
Phys. Rev. Lett. \textbf{106}, 230602 (2011).

\bibitem{Saito2011}
K. Saito, G. Benenti, G. Casati, and T. Prosen,
Phys. Rev. B \textbf{84}, 201306(R) (2011).

\bibitem{Sanchez2011}
D. S\'anchez and L. Serra,
Phys. Rev. B \textbf{84}, 201307(R) (2011).

\bibitem{Horvat2012}
M. Horvat, T. Prosen, G. Benenti, and G. Casati,
Phys. Rev. E \textbf{86}, 052102 (2012).

\bibitem{Vinitha2013}
V. Balachandran, G. Benenti, and G. Casati,
Phys. Rev. B \textbf{87}, 165419 (2013).

\bibitem{Brandner2013a}
K. Brandner, K. Saito, and U. Seifert,
Phys. Rev. Lett. \textbf{110}, 070603 (2013).

\bibitem{Brandner2013b}
K. Brandner and U. Seifert,
New J. Phys. \textbf{15}, 105003 (2013).

\bibitem{Brandner2015}
K. Brandner and U. Seifert,
Phys. Rev. E \textbf{91}, 012121 (2015).

\bibitem{Yamamoto2016}
K. Yamamoto, O. Entin-Wohlman, A. Aharony, and N. Hatano,
Phys. Rev. B \textbf{94}, 121402(R) (2016).

\bibitem{Luo2020}
R. Luo, G. Benenti, G. Casati, and J. Wang,
Phys. Rev. Res. \textbf{2}, 022009(R) (2020).

\bibitem{Saito2016}
N. Shiraishi, K. Saito, and H. Tasaki,
Phys. Rev. Lett. \textbf{117}, 190601 (2016).

\bibitem{Allahverdyan2013}
A. E. Allahverdyan, K. V. Hovhannisyan, A. V. Melkikh, and S. G. Gevorkian,
Phys. Rev. Lett. \textbf{111}, 050601 (2013).

\bibitem{Shiraishi2015}
N. Shiraishi,
Phys. Rev. E \textbf{92}, 050101 (2015).

\bibitem{Campisi2016}
M. Campisi and R. Fazio,
Nat. Commun. \textbf{7}, 11895 (2016).

\bibitem{Koning2016}
J. Koning and J. O. Indekeu,
Eur. Phys. J. B \textbf{89}, 248 (2016).

\bibitem{Polettini2017}
M. Polettini and M. Esposito,
Europhys. Lett. \textbf{118}, 40003 (2017).

\bibitem{Lee2017}
J. S. Lee and H. Park,
Sci. Rep. \textbf{7}, 10725 (2017).

\bibitem{Holubev2017}
V. Holubec and A. Ryabov,
Phys. Rev. E \textbf{96}, 030102(R) (2017).

\bibitem{Barato2015}
A. C. Barato and U. Seifert,
Phys. Rev. Lett. \textbf{114}, 158101 (2015).

\bibitem{Gingrich2016}
T. R. Gingrich, J. M. Horowitz, N. Perunov, and J. L. England,
Phys. Rev. Lett. \textbf{116}, 120601 (2016).

\bibitem{Seifert2019}
U. Seifert, Ann. Rev. Cond. Mat. Phys. \textbf{10}, 171 (2019).

\bibitem{Horowitz2020}
J. M. Horowitz and T. R. Gingrich, Nature Physics \textbf{16},
15 (2020).

\bibitem{Pietzonka2018}
P. Pietzonka and U. Seifert,
Phys. Rev. Lett. \textbf{120}, 190602 (2018).

\bibitem{footnote_DeltaP}
Since $P (t)$ converges for $t \to \infty$ to $P$ as
$1/\sqrt{t}$, the factor $t$ in (\ref{eq:DeltaP}) is needed
to obtain a finite limit for $\Delta_P$.

\bibitem{Benenti2013}
G. Benenti, G. Casati, and J. Wang,
Phys. Rev. Lett. \textbf{110}, 070604 (2013).

\bibitem{Benenti2014}
G. Benenti, G. Casati, and C. Mej\'{\i}a-Monasterio,
New J. Phys. \textbf{16}, 015014 (2014).

\bibitem{Chen2015}
S. Chen, J. Wang, G. Casati, and G. Benenti,
Phys. Rev. E \textbf{92}, 032139 (2015).

\bibitem{footnote_modes}
For the sake of simplicity we consider here systems with a single transverse mode.
Considering $\mathcal{N}$ transverse modes does not change our results for the
trade-off parameter $\mathcal{Q}$.

\bibitem{Whitney2014}
R. S. Whitney, Phys. Rev. Lett. \textbf{112}, 130601 (2014).

\bibitem{Whitney2015}
R. S. Whitney, Phys. Rev. B \textbf{91}, 115425 (2015).

\bibitem{mahan}
G. D. Mahan and J. O. Sofo,
Proc. Natl. Acad. Sci. USA \textbf{93}, 7436 (1996)

\bibitem{linke1}
T. E. Humphrey, R. Newbury, R. P. Taylor, and H. Linke,
Phys. Rev. Lett. \textbf{89}, 116801 (2002).

\bibitem{linke2}
T. E. Humphrey and H. Linke,
Phys. Rev. Lett. \textbf{94}, 096601 (2005).

\bibitem{Levitov1993}
L. S. Levitov and G. B. Lesovik, JETP Lett. \textbf{58}, 230 (1993).

\bibitem{Nazarov2009}
Y. V. Nazarov and Y. M. Blanter,
\textit{Quantum Transport: Introduction to Nanoscience}
(Cambridge University Press, Cambridge, 2009).

\bibitem{Segal2019}
J. Liu and D. Segal,
Phys. Rev. E \textbf{99}, 062141 (2019).

\bibitem{Luo2018}
R. Luo, G. Benenti, G. Casati, and J. Wang,
Phys. Rev. Lett. \textbf{121}, 080602 (2018).

\bibitem{supp}
See Supplemental Material for details on the derivation of
Eq.~(\ref{eq:expandQ}), which includes Refs.~\cite{Whitney2015,
Segal2019, Luo2018, Saito2010, Gaspard2013, Brandner2018}.

\bibitem{Saito2010}
K. Saito, G. Benenti, and G. Casati,
Chem. Phys. \textbf{375}, 508 (2010).

\bibitem{Gaspard2013}
P. Gaspard,
New J. Phys. \textbf{15}, 115014 (2013).

\bibitem{Brandner2018}
K. Brandner, T. Hanazato, and K. Saito,
Phys. Rev. Lett. \textbf{120}, 090601 (2018).

\bibitem{reservoir}
C. Mej\'{\i}a-Monasterio, H. Larralde, and F. Leyvraz,
Phys. Rev. Lett. \textbf{86}, 5417 (2001);
H. Larralde, F. Leyvraz, and C. Mej\'{\i}a-Monasterio,
J. Stat. Phys. \textbf{113}, 197 (2003).

\bibitem{Callen}
H. B. Callen,
\textit{Thermodynamics and an Introduction to Thermostatics}
(2nd ed.)
(John Wiley \& Sons, New York, 1985).

\bibitem{Groot}
S. R. de Groot and  P. Mazur,
\textit{Nonequilibrium Thermodynamics}
(North-Holland, Amsterdam, 1962).

\bibitem{Benenti2017}
G. Benenti, G. Casati, K. Saito, and R. S. Whitney,
Phys. Rep. \textbf{694}, 1 (2017).

\bibitem{Macieszczak2018}
K. Macieszczak, K. Brandner, and J. P. Garrahan,
Phys. rev. Lett. \textbf{121}, 130601 (2018).

\bibitem{Brantut2013}
J.-P. Brantut, C. Grenier, J. Meineke, D. Stadler, S. Krinner,
C. Kollath, T. Esslinger, and A. Georges,
Science \textbf{342}, 713 (2013).

\bibitem{Husmann2018}
D. Husmann, M. Lebrat, S. H\"{a}usler, J.-P. Brantut, L. Corman,
and T. Esslinger,
PNAS \textbf{115}, 8563 (2018).

\end{thebibliography}
\end{document}